\documentclass[12pt]{article}

\usepackage{amsmath}
\usepackage{graphicx}
\usepackage[font=small,margin=10pt,labelfont=bf]{caption}
\usepackage{subcaption}
\usepackage{array}                    
\usepackage[linkbordercolor={0.28 0.28 0.43}]{hyperref}
\usepackage{multirow}
\usepackage[table]{xcolor}
\usepackage{authblk} 

\begin{document}

\title{Revisiting Routh's Classification of the Heavy Symmetric Top: A Correction}

\author[1]{Vedat Tanr{\i}verdi\thanks{Email: \href{mailto:tanriverdivedat@googlemail.com}{tanriverdivedat@googlemail.com} \\ ORCID: \href{https://orcid.org/0000-0003-0225-2183}{0000-0003-0225-2183}}}
\affil[1]{Ankara Science University, 06570 Ankara, Turkey}

\date{\today}

\maketitle

\begin{abstract}
The classification of the motion of a heavy symmetric top when conserved angular momenta are different is studied by considering modern method and Routh's geometrical method.
It is shown that Routh's method provides correct classification for four cases; however, it fails in two cases.

\vspace{0.3cm}
\end{abstract}

\vspace{0.5cm}

\section{Introduction}

The symmetric top is known since ancient times \cite{brecher2014topology,featonby2010dare}.
Plato describes the motion of the upright top in the Republic as: "... tops stand still as a whole and move at the same time, when they spin around with the point fixed in the same place ..."\cite{Plato}. 
However, its scientific study starts with Euler's derivations of rigid body equations \cite{Euler1752, Euler1765}.
Many important scientists have studied this problem like Lagrange \cite{lagrange_anme}, Poinsot \cite{Goldstein}, Kovalevskaya \cite{Kovalevskaya}, Klein \& Sommerfeld \cite{KleinSommerfeld}.

While these authors established the foundational equations, a comprehensive classification of the top's motion based on energy and conserved angular momenta was notably provided by Routh \cite{Routh}.
The first edition of his work was published in 1860, and the final edition was published in 1905. 
It is one of the detailed works on the classification of the motion.
None of the other scientific works provide a classification of the motion of the heavy symmetric top similar to Routh's work, i.e. considering conserved angular momenta and energy.
Though Klein \& Summerfeld's work is one of the extensive ones and considers different types of motion, it also does not provide a classification by using energy and conserved angular momenta.

Most of the classical mechanics books provide explanations related to the upright top, regular precession and three motions when conserved angular momentum around the symmetry axis is greater than the conserved angular momentum around vertical axis without providing energy relation \cite{Goldstein, LandauLifshitz, Arnold, Corinaldesi, MarionThornton, Arya, Symon}. 

In this work, the classification of the motion will be studied with modern methods when conserved angular momenta are different.
For comparison, Routh's method will also be considered.
In the second section, theoretical background will be considered, and in the third section these two classifications will be studied.
Finally the situation will be concluded.

\section{Theoretical Background}

In this section, the constants of motion for a heavy symmetric top will be provided with modern method, whose parameters are available in recent classical mechanics books e.g. \cite{Goldstein}, and geometric method used by Routh \cite{Routh}.

For the classification of motion, in general, the motion of the symmetry axis is considered and it can be shown on a unit sphere, i.e. shapes for the locus of the symmetry axis. This diagram is mainly determined by angular velocities $\dot \theta $ and $\dot \phi$.

\subsection{Modern Method}

The motion of a heavy symmetric top can be described in terms of Euler angles. $\theta$ is the angle between vertical and body symmetry axis, $\phi$ is the angle related to rotation of the top around vertical direction, and $\psi$ is the angle related to the spin of the top. 

For the motion of a heavy symmetric top, energy and two components of angular momentum are conserved. These can be written as \cite{Goldstein}
\begin{equation}
    E=\frac{I_x}{2}(\dot \theta ^2 + \dot \phi ^2 \sin^2 \theta)+\frac{I_z}{2}(\dot \psi+\dot \phi \cos \theta)^2+M g l \cos \theta.
    \label{eq:enrgy}
\end{equation}
\begin{eqnarray}
    p_{\psi}&=&I_z(\dot \psi+\dot \phi \cos \theta)=I_x a, \nonumber \\
    p_{\phi}&=&I_x \dot\phi \sin^2 \theta +I_z (\dot \psi+\dot \phi \cos \theta)\cos \theta=I_x b .
    \label{eq:angmom}
\end{eqnarray}

By using the constancy of $p_{\psi}$, it is possible to define another constant corresponding to the energy in terms of $a$ and $b$ as \cite{Goldstein}
\begin{equation}
    E'=\frac{I_x}{2}\dot \theta ^2 +\frac{I_x}{2} \frac{(b-a \cos \theta)^2}{\sin^2 \theta}+M g l \cos \theta. \label{eq:eprime}
\end{equation}

From $E'$, an effective potential can be defined as \cite{Goldstein}
\begin{equation}
    V'(\theta)= \frac{I_x}{2}\frac{(b-a \cos \theta)^2}{\sin^2 \theta}+Mgl \cos \theta.
    \label{eq:ueff}
\end{equation}
The form of the effective potential is similar to well when $|a|\ne|b|$ and intersections of $E'$ with the effective potential give turning points for $\theta$.

$\dot \phi$ can be obtained in terms of $a$ and $b$ as a function of $\theta$ \cite{Goldstein}
\begin{equation}
    \dot \phi(\theta)=\frac{b - a \cos \theta}{\sin^2 \theta}. \label{eq:phidot}
\end{equation}

When $V'=Mglb/a$, from the effective potential one can arrive at the relation 
\begin{equation}
    \dot \phi^2= \frac{2Mgl}{I_x a} \dot \phi.
    \label{eq:mglboadotphi}
\end{equation}
This shows that when $V'=Mglb/a$, $\dot \phi$ is equal to either $0$ or $2Mgl /(I_x a)$. And, this is one of the important factors determining shapes for the locus of the symmetry axis.
It can be seen from equation (\ref{eq:phidot}) that $\dot \phi=0$ is only possible when $|a|>|b|$. 
Hence, when $|b|>|a|$, $\dot \phi$ always has the same sign and shapes for the locus of the symmetry axis are always similar.

\subsection{Routh's Geometrical Method}

In this part, we will give the relations between quantities used in this work and parameters used by Routh \cite{Routh}.
To avoid confusion between the parameters, we will designate parameters and quantities used by Routh in square parentheses.

The correspondences between symbols representing moments of inertia are $[A]=I_x$, $[B]=I_y$ and $[C]=I_z$, for symmetric top $[A]=[B]$.
The relations between angular velocities are as follows:
\begin{eqnarray}
    [w_3] &=& [n] =\dot \psi+\dot \phi \cos \theta, \\
    {[}-w_1] &=& [\dot \Psi \sin \theta]= \dot \phi \sin \theta, \\
    {[}w_2] &=& \dot \theta.
\end{eqnarray}

In equation (2) of Art. 200, Routh has given the relation $[-A w_1 \sin \theta+ C n \cos \theta]=[E]$ which can be written as $I_x \dot \phi \sin^2 \theta+I_z (\dot \psi+\dot \phi \cos \theta)\cos \theta$, and this is equal to $p_{\phi}=I_x b$, then $[E]=I_x b$.

There are four parameters in Routh's work to determine different types of motion (Art. 204): $[l]$, $[c]$, $[a]$ and $[OQ]$.
The length $[l]=[A/h]$ used by Routh can be written in terms of quantities used in this work as 
\begin{equation}
[l]=I_x/l.
\end{equation}
The second parameter is given by $[c]=[l^2\left( C(w_1^2+w_2^2)+2 g h w_1 \sin \theta/ n\right)/(2Cg)]$ and it can be obtained as 
\begin{equation}
[c]=\frac {I_x}{l} \frac {1}{Mgl} \left(E'-Mgl \frac{b}{a}\right),
\end{equation}
which provides a relation between $E'$ and $Mgl \frac{b}{a}$.
The third parameter is given by $[a]=[E l/(Cn)]$ which can be obtained as 
\begin{equation}
[a]= \frac{I_x}{l} \frac{b}{ a},
\end{equation}
which correspons to the ratio of conserved angular momenta.

The next parameter gives a bit of a strange result in terms of the quantities used in this work.
From the last lines of Art. 204 b, it can be inferred that $[OQ^2]$ is equivalent to $[(a+2c)^2-8plc]$ where $[p]=[C^2 n^2/(4 g h^2 l)]$ which can be seen from the explanation after equation (2) in Art. 204.
Then, it is possible to write $[p]=I_x a^2/(4 Mgl)$.
Hence
\begin{equation}
    [OQ^2]=\left( \frac{I_x}{l}\right)^2 \Bigg\{  \frac{b^2}{a^2}+\frac{4}{(Mgl)^2} \left( E'-Mgl \frac{b}{a}\right)\left(E'-\frac{I_x a^2}{2}\right) \Bigg\}.
    \label{paraoq}
\end{equation}

\section{Different types of motion}

In Routh's work, different types of motion are considered in three different parts by considering the parameter $[c]$: 
In the first ($[c]>0$) and third ($[c]=0$) parts, the relation between $[l]$ and $[a]$ is considered, and in the second part ($[c]<0$), the relation between $[l]$ and $[OQ]$ is considered.
Here, the first and the third part will be considered before the second part. 
$[a]$, $a$ and $b$ will be coonsidered as positive for the following parts. For negative values of $a$ and $b$, one can see \cite{tanriverdi_abdif}.

For the numerical calculations, the following values are used $I_x=I_y=8.44\times 10^{-5}\,\mathrm{kg\,m^2}$, $I_z=7.23\times 10^{-5}\,\mathrm{kg\,m^2}$ and $Mgl=0.215\,\mathrm{J}$ \cite{Cross2013, Tanriverdi2020b}.
For numerical solutions, the method explained in \cite{Tanriverdi2019} is used.

\subsection{$[c]>0$ or $E'>Mgl\frac{b}{a}$}

In this case, when $a>b$, $\dot \phi$ can change sign. On the other hand, when $b>a$, $\dot \phi$ cannot change sign.
Then, the heavy symmetric top makes looping motion when $a>b$, and it continues to precess at the same direction always when $b>a$.

\begin{figure}[!h]
       \begin{center}
           \begin{subfigure}{0.45\textwidth}
               \centering
                   \includegraphics[width=5.0cm]{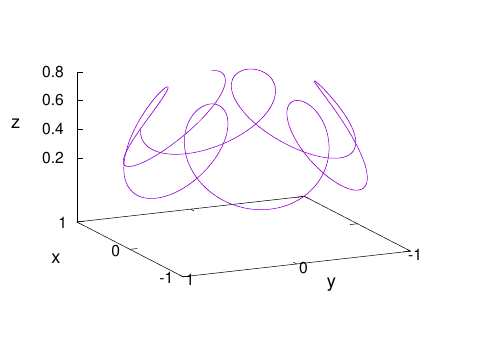}
		   \caption{$a>b$ and $E'>Mgl\frac{b}{a}$}
           \end{subfigure}
               \begin{subfigure}{0.45\textwidth}
            \centering
                       \includegraphics[width=5.0cm]{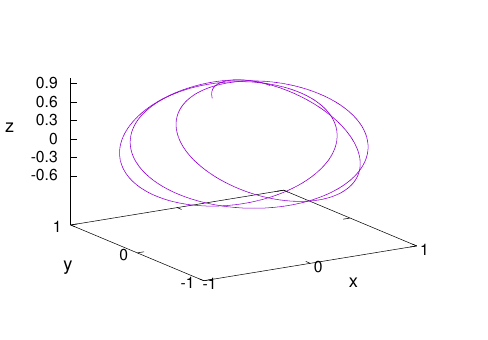}
			\caption{$b>a$ and $E'>Mgl\frac{b}{a}$}
               \end{subfigure}
	       \caption{ Shapes for the locus of the symmetry axis on the unit sphere. 
            (a) Initial values are $\theta_0=0.648\,\mathrm{rad}$, $\dot \theta_0=0$, $\dot \phi_0=-43.1\,\mathrm{rad\,s^{-1}}$ and $\dot \psi_0= 174\,\mathrm{rad\,s^{-1}}$, and constants are $a=120\,\mathrm{rad\,s^{-1}}$, $b=80.0\,\mathrm{rad\,s^{-1}}$, $E'=0.200\,\mathrm{J}$, and $Mgl b/a=0.143\,\mathrm{J}$.
	    The animated plot is available at \href{https://youtu.be/xg7PFi33GGY}{https://youtu.be/xg7PFi33GGY}.
		(b) Initial values are $\theta_0=0.251\,\mathrm{rad}$, $\dot \theta_0=0$, $\dot \phi_0=188\,\mathrm{rad\,s^{-1}}$ and $\dot \psi_0=-124\,\mathrm{rad\,s^{-1}}$, and constants are $a=50.0\,\mathrm{rad\,s^{-1}}$, $b=60.0\,\mathrm{rad\,s^{-1}}$, $E'=0.300\,\mathrm{J}$, and $Mgl b/a=0.258\,\mathrm{J}$.
                The animated plot is available at \href{https://youtu.be/CZcHT6N4OKU}{https://youtu.be/CZcHT6N4OKU}.
           }
       \label{fig:epbtmglboa}
       \end{center}
\end{figure}

Projections of shapes for the locus of the symmetry axis on horizontal plane can be seen from figure \ref{fig:epbtmglboa}.
$b=a$ case is studied previously, and when $E'>Mgl\frac{b}{a}$, the top can make "motion of a fast top" or "motion over the bump" \cite{Tanriverdi2020} which are similar to the figure provided by Routh.

In Routh's work, the motion is classified by using the relation between $[l]$ and $[a]$.
$[l]>[a]$ corresponds to $a>b$, $[l]<[a]$ corresponds to $b>a$, and $[l]=[a]$ corresponds to $a=b$.
It can be seen that Routh's classification and considerations in this work are consistent.

\subsection{$[c]=0$ or $E'=Mgl\frac{b}{a}$}

In this case, when $E'=Mgl\frac{b}{a}$; if $a>b$, $\dot \phi$ can acquire zero at one of the turning angles of $\theta$, but if $b>a$, $\dot \phi$ cannot be zero.
Then, the top makes motion with cusps when $a>b$, and again it continues to precess always in the same direction when $b>a$.

\begin{figure}[!h]
       \begin{center}
       \begin{subfigure}{0.45\textwidth}
               \centering
           \includegraphics[width=5.0cm]{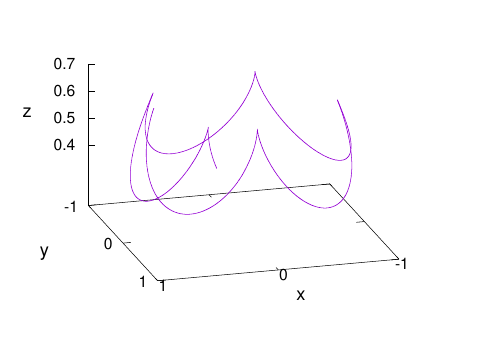}
		\caption{$a>b$ and $E'=Mgl\frac{b}{a}$}
       \end{subfigure}
       \begin{subfigure}{0.45\textwidth}
               \centering
               \includegraphics[width=5.0cm]{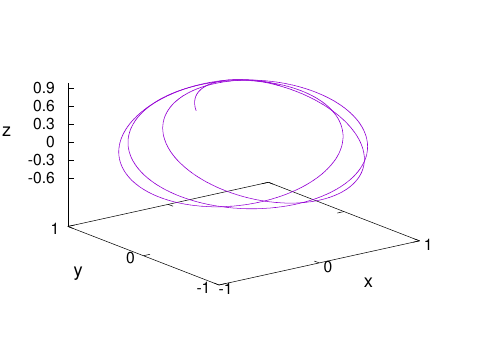}
		\caption{$b>a$ and $E'=Mgl\frac{b}{a}$}
       \end{subfigure}

           \caption{ Shapes for the locus of the symmetry axis on the unit sphere. 
	       (a) Initial values are $\theta_0=0.841\,\mathrm{rad}$, $\dot \theta_0=0$, $\dot \phi_0= 0$ and $\dot \psi_0= 140\,\mathrm{rad\,s^{-1}}$, and constants are $a=120\,\mathrm{rad\,s^{-1}}$, $b=80.0\,\mathrm{rad\,s^{-1}}$, $E'=0.143\,\mathrm{J}$ and $Mgl b/a=0.143\,\mathrm{J}$.
           The animated plot is available at \href{https://youtu.be/TueiP-7EDEE}{https://youtu.be/TueiP-7EDEE}.
                (b) Initial values are $\theta_0=0.371\,\mathrm{rad}$, $\dot \theta_0=0$, $\dot \phi_0=102\,\mathrm{rad\,s^{-1}}$ and $\dot \psi_0=-36.6\,\mathrm{rad\,s^{-1}}$, and constants are $a=50.0\,\mathrm{rad\,s^{-1}}$, $b=60.0\,\mathrm{rad\,s^{-1}}$, $E'=0.258\,\mathrm{J}$ and $Mgl b/a=0.258\,\mathrm{J}$.
                The animated plot is available at \href{https://youtu.be/A5tXLg_wCJo}{https://youtu.be/A5tXLg\_wCJo}.
           }
       \label{fig:epeqmglboa}
       \end{center}
\end{figure}

Projections of shapes for the locus of the symmetry axis on horizontal plane can be seen from figure \ref{fig:epeqmglboa}.
Again from the previous study\cite{Tanriverdi2020}, for a fast top ($a>\sqrt{4Mgl/I_x}$) the top stays at the upwrigt position, for a slow top ($a<\sqrt{4Mgl/I_x}$) it makes spiralling motion as long as $\theta>0$.
Routh has provided the shapes for the locus of the symmetry axis only for spiralling motion.

In Routh's work again the relation between $[l]$ and $[a]$ is considered to classify motion.
The figures in Routh's work are similar to given ones in figure \ref{fig:epeqmglboa}.

\subsection{$[c]<0$ or $E'<Mgl\frac{b}{a}$}

\begin{figure}[!h]
       \begin{center}
       \begin{subfigure}{0.45\textwidth}
               \centering
           \includegraphics[width=5.0cm]{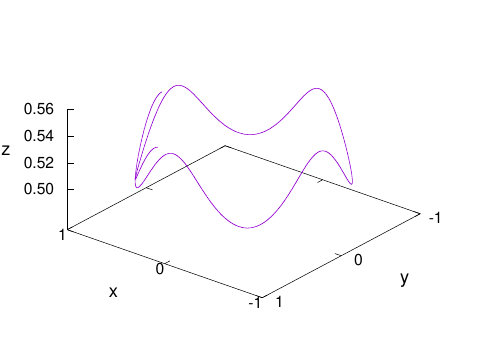}
		\caption{$a>b$, $E'<Mgl\frac{b}{a}$ and $1>\eta$}
       \end{subfigure}
       \begin{subfigure}{0.45\textwidth}
               \centering
               \includegraphics[width=5.0cm]{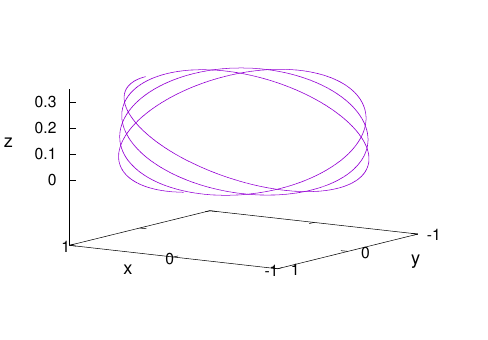}
		\caption{$b>a$, $E'<Mgl\frac{b}{a}$ and $1<\eta$}
       \end{subfigure}
	       \caption{ Shapes for the locus of the symmetry axis on the unit sphere. 
	       (a) Initial values are $\theta_0=0.993\,\mathrm{rad}$, $\dot \theta_0=0$, $\dot \phi_0= 20.6\,\mathrm{rad\,s^{-1}}$ and $\dot \psi_0= 129\,\mathrm{rad\,s^{-1}}$, and constants are $a=120\,\mathrm{rad\,s^{-1}}$, $b=80.0\,\mathrm{rad\,s^{-1}}$, $E'=0.130\,\mathrm{J}$, $Mgl b/a=0.143\,\mathrm{J}$, and $\eta=0.996$.
           The animated plot is available at \href{https://youtu.be/PjJdCahiPEQ}{https://youtu.be/PjJdCahiPEQ}.
                (b) Initial values are $\theta_0=1.23\,\mathrm{rad}$, $\dot \theta_0=0$, $\dot \phi_0=60.0\,\mathrm{rad\,s^{-1}}$ and $\dot \psi_0= 38.3\,\mathrm{rad\,s^{-1}}$, and constants are $a=50.0\,\mathrm{rad\,s^{-1}}$, $b=70.0\,\mathrm{rad\,s^{-1}}$, $E'=0.207\,\mathrm{J}$, $Mgl b/a=0.301\,\mathrm{J}$, and $\eta=1.13$.
                The animated plot is available at \href{https://youtu.be/gJbHvj-fb5s}{https://youtu.be/gJbHvj-fb5s}.
           }
       \label{fig:epstmglboa}
       \end{center}
\end{figure}

In this case, $\dot \phi$ cannot become zero since $E'<Mgl\frac{b}{a}$. 
Then, the top makes wavy precession when $a>b$, and again it continues to precess at the same direction always when $b>a$.
Projections of shapes for the locus of the symmetry axis on horizontal plane for the first two cases can be seen from figure \ref{fig:epstmglboa}. 

Again from the previous study \cite{Tanriverdi2020}, for a slow top ($a<\sqrt{4Mgl/I_x}$) the top makes "motion for slow top when $b=a$ and $V'_{min} < E' < Mgl$" or regular precession.
There is not any mention of these in Routh's classification when $[c]<0$.

\begin{figure}[!h]
       \begin{center}
       \begin{subfigure}{0.45\textwidth}
               \centering
           \includegraphics[width=5.0cm]{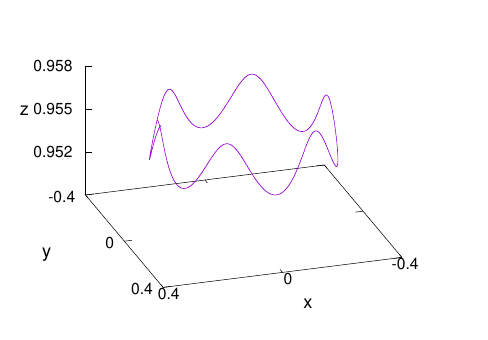}
		\caption{$a>b$, $E'<Mgl\frac{b}{a}$ and $1<\eta$}
       \end{subfigure}
       \begin{subfigure}{0.45\textwidth}
               \centering
               \includegraphics[width=5.0cm]{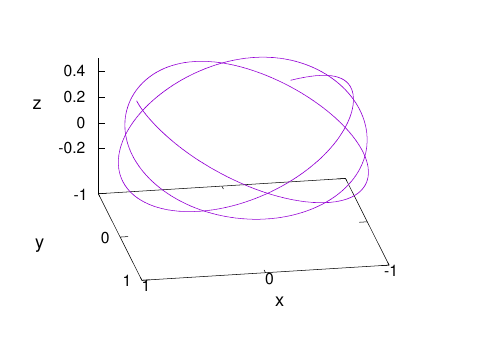}
		\caption{$b>a$, $E'<Mgl\frac{b}{a}$ and $1>\eta$}
       \end{subfigure}

           \caption{ Shapes for the locus of the symmetry axis on the unit sphere. 
	       (a) Initial values are $\theta_0=0.294\,\mathrm{rad}$, $\dot \theta_0=0$, $\dot \phi_0= 17.0\,\mathrm{rad\,s^{-1}}$ and $\dot \psi_0= 159\,\mathrm{rad\,s^{-1}}$, and constants are $a=150\,\mathrm{rad\,s^{-1}}$, $b=145\,\mathrm{rad\,s^{-1}}$, $E'=0.207\,\mathrm{J}$, $Mgl b/a=0.208\,\mathrm{J}$, and $\eta=1.000018 $.
           The animated plot is available at \href{https://youtu.be/azOP55Lgyyc}{https://youtu.be/azOP55Lgyyc}. 
                (b) Initial values are $\theta_0=1.05\,\mathrm{rad}$, $\dot \theta_0=0$, $\dot \phi_0=43.4\,\mathrm{rad\,s^{-1}}$ and $\dot \psi_0= 42.6\,\mathrm{rad\,s^{-1}}$, and constants are $a=55.0\,\mathrm{rad\,s^{-1}}$, $b=60.0\,\mathrm{rad\,s^{-1}}$, $E'=0.167\,\mathrm{J}$, $Mgl b/a=0.234\,\mathrm{J}$, and $\eta= 0.960605$.
                The animated plot is available at \href{https://youtu.be/a1Edvb34Ao0}{https://youtu.be/a1Edvb34Ao0}.
           }
       \label{fig:epstmglboa_routh}
       \end{center}
\end{figure}

In Routh's work, for this case, instead of $[a]$, $[OQ]$ is used.
It is possible to define a parameter $\eta$ by considering the relation between $[l]$ and $[OQ]$ as
\begin{equation}
\eta=\frac{b^2}{a^2}+\frac{4}{(Mgl)^2} \left( E'-Mgl \frac{b}{a}\right)\left(E'-\frac{I_x a^2}{2}\right)
\end{equation}
Then, the relation between $[l^2]>[OQ^2]$ can be written as $1>\eta$, and $[l^2]<[OQ^2]$ can be written as $1<\eta$.
From these relations, it is possible to obtain similar results shown in figure \ref{fig:epstmglboa}.
However, it is also possible to obtain results opposite to these, which can be seen in figure \ref{fig:epstmglboa_routh}.

It is clear that motion types given in Routh's work are consistent with figure \ref{fig:epstmglboa}, and they are not consistent with figure \ref{fig:epstmglboa_routh}.

\section{Conclusion}

It can be seen that Routh's classification of the motion of a heavy symmetric top is true for the cases $[c]>0$ ($E'>Mgl b/a$) and $[c]=0$ ($E'=Mgl b/a$), and it yields inconsistencies for the case $[c]<0$ ($E'< Mgl b/a$).
Had Routh used $[a]$ instead of $[OQ]$, his classification would have been correct.

\section*{Acknowledgments}

The author thanks the anonymous referee for their valuable feedback, and acknowledges the use of the AI assistant Gemini for language polishing and manuscript formatting.

The author declares that no funding, grants, or other financial support was received for the research, authorship, or publication of this article.

The author declares that he has no potential conflicts of interest or competing financial/personal relationships that could have appeared to influence the work reported in this paper.

The datasets generated and analyzed during the current study, along with the custom computer code used for simulation are available from the corresponding author upon reasonable request.

\end{document}